\documentclass[pra,aps,amsmath,amssymb,showkeys]{revtex4}
\newcommand{\pen}{\openone}
\newcommand{\bro}{{\hat{\rho}}}
\newcommand{\amh}{{\hat{A}}}
\newcommand{\bmh}{{\hat{B}}}
\newcommand{\mmh}{{\hat{M}}}
\newcommand{\nnh}{{\hat{N}}}

\newcommand{\skx}{{\hat{k}}}
\newcommand{\sqx}{{\hat{q}}}
\newcommand{\spx}{{\hat{p}}}
\newcommand{\sax}{{\hat{x}}}

\newcommand{\itt}{{\mathtt{i}}}
\newcommand{\xdif}{{\mathrm{d}}}
\newcommand{\cld}{{\mathcal{D}}}
\newcommand{\clk}{{\mathcal{K}}}
\newcommand{\clm}{{\mathcal{M}}}
\newcommand{\cln}{{\mathcal{N}}}
\newcommand{\clq}{{\mathcal{Q}}}
\newcommand{\clx}{{\mathcal{X}}}
\newcommand{\Tr}{{\mathrm{Tr}}}
\begin{document}
\clearpage
\preprint{}

\title{On entropic uncertainty relations in the presence of a minimal length}

\author{Alexey E. Rastegin}

\affiliation{Department of Theoretical Physics, Irkutsk State University,
Gagarin Bv. 20, Irkutsk 664003, Russia}

\begin{abstract}
Entropic uncertainty relations for the position and momentum
within the generalized uncertainty principle are examined. Studies
of this principle are motivated by the existence of a minimal
observable length. Then the position and momentum operators
satisfy the modified commutation relation, for which more than one
algebraic representation is known. One of them is described by
auxiliary momentum so that the momentum and coordinate wave
functions are connected by the Fourier transform. However, the
probability density functions of the physically true and auxiliary
momenta are different. As the corresponding entropies differ,
known entropic uncertainty relations are changed. Using
differential Shannon entropies, we give a state-dependent
formulation with correction term. State-independent uncertainty
relations are obtained in terms of the R\'{e}nyi entropies and the
Tsallis entropies with binning. Such relations allow one to take
into account a finiteness of measurement resolution.
\end{abstract}


\maketitle

\pagenumbering{arabic}
\setcounter{page}{1}

\section{Introduction}\label{sec1}

One of fundamental problems of modern physics is to describe the
gravitation at the quantum level \cite{rovelli04}. Today,
theoretical efforts are focused on unifying all fundamental
interactions into a single theoretical framework. The existence
of a minimal observable length has long been suggested due to such
studies \cite{garay95,hossen13}. It should lead to an effective
cutoff in the ultraviolet \cite{kempf97}. String-theoretic
arguments also maintain a minimal length effectively in the form
of a minimal position uncertainty. There are proposals to
investigate observable effects of the minimal length, including
astronomical observations \cite{ellis98,piran07} and experimental
schemes feasible within current technology \cite{brukner12,ffm13}.
The authors of \cite{akhv15,bousso16,hhbf16} discussed
measurements in which we may be able to prove effects of quantum
gravity. The role of quantum decoherence in modern particle
experiments is emphasized in \cite{alok15}.

The uncertainty principle \cite{heisenberg} is well known among
scientific achievements inspired by the discovery of quanta.
Discussion of Heisenberg \cite{heisenberg} was rather qualitative
in character. There is no general consensus concerning scope and
validity of the uncertainty principle \cite{lahti}. One of topics
questioned concerns proper forms of the uncertainty principle
beyond the standard quantum mechanics. In many models of quantum
gravity and string theory, the Planck length
$\ell_{P}=\sqrt{G\hbar/c^{3}}\approx1.616\times10^{-35}$ m plays a
crucial role. It seems that the very concept of space-time changes
its meaning below the Planck scale \cite{amati89}. Processes
around the Planck energy may depend on physical effects of virtual
black holes \cite{faizal12}. At this scale, Heisenberg's principle
is assumed to be converted into the generalized uncertainty
principle (GUP) \cite{scard99,bombi08}. Of course, this
modification should be consistent with the existence of a minimal
observable length. The GUP issue has many aspects that are
currently the subject of active researches
\cite{tawfik13,diab14,faizalm15,faizalk15,gdf15,faizal16,mfza16,chwen16}.

To get a quantum model with a nonzero minimal uncertainty in
position, the commutation relation for position and momentum is
modified. Consequences of deformed forms of the commutation
relation have attracted much attention in recent years, though
their connections with the real world are an open question. In the
context of non-relativistic quantum mechanics, the corresponding
formalism was developed in \cite{kempf95}. The author of
\cite{pedram12} proposed another approach to representation of the
position and momentum operators. Being physically equivalent to
the representation of \cite{kempf95}, it differs in some formal
aspects. The above approaches used the momentum representation for
dealing with the Schr\"{o}dinger equation in the GUP case. It was
recently shown that the position representation of this case is
possible with quasi-nonlinear evolution equation \cite{lrud16}.
Path integral quantization corresponding to the deformed algebra
was examined in \cite{pmfa15}. The authors of \cite{acfms16}
thoroughly considered physical assumptions under which the used
modifications actually imply a minimal length.

The first explicit derivation of uncertainty relations was given
by Kennard \cite{kennard}: product of the standard deviations of
the position and momentum operators cannot be less than the
constant equal to $\hbar/2$. Robertson \cite{robert} extended this
approach to arbitrary pair of observables. The authors of
\cite{IZBB12a,IZBB12b,IZBB13} discussed uncertainty relations for
photons. Due to lack of well-defined position operator of a
photon, their method focuses on the electromagnetic energy
distribution in space. Robertson's formulation was later
criticized for several reasons \cite{deutsch,maass}. Instead, an
entropic formulation of the uncertainty principle has been
proposed and motivated \cite{deutsch,kraus,maass,krishna}.
Entropic uncertainty relations are currently the subject of active
research (see, e.g., the reviews \cite{ww10,brud11,cbtw15} and
references therein). In finite dimensions, uncertainty relations
can be posed in terms of very wide class of entropic functions
\cite{zbp2014}. Other approaches are based on the sum of variances
\cite{huang12,mpati14}, majorization relations
\cite{prz13,fgg13,rpz14,lrud15,arkz16} and the technique of
effective anti-commutators \cite{ktw14}.

Entropic functions provide a clean and flexible tool for
characterizing uncertainties in quantum measurements. The first
entropic uncertainty relation for the position-momentum pair was
derived by Hirschman \cite{hirs} and later improved in
\cite{beck,birula1}. Entropic uncertainty relations in
multi-dimensional spaces were given in \cite{huang11}. In the
information-theoretic framework, uncertainty relations in the
presence of a quantum memory were formulated
\cite{BCCRR10,fbtsc14}. Recently, the generalized uncertainty
principle has been analyzed within the entropic approach
\cite{pedram16,khww16}. These formulations are expressed in terms
of the Shannon entropies. On the other hand, more general forms of
entropic functions have found use in quantifying uncertainties.
Utilities of entropic bounds with a parametric dependence were
first emphasized in \cite{maass}. In particular, such bounds may
allow us to find more exactly the domain of acceptable values for
unknown probabilities with respect to known ones.

The aim of the present work is to examine the generalized
uncertainty principle in terms of entropic functions of both the
R\'{e}nyi and Tsallis types. We will focus on those questions that
were not addressed in this context previously. The paper is
organized as follows. In Section \ref{sec2}, we review preliminary
material including details of the used representation of the
position and momentum operators. A state-dependent formulation of
the generalized uncertainty principle is obtained in Section
\ref{sec3} with the use of Shannon entropies. In Section
\ref{sec4}, we obtain state-independent uncertainty relations in
terms of the R\'{e}nyi entropies and, with appropriate binning,
the Tsallis entropies. In Section \ref{sec5}, we conclude the
paper with a brief summary of results.

\section{Preliminaries}\label{sec2}

In this section, we recall the required material and describe the
notation. For convenience, we will use the wavenumber operator
$\skx$ instead of the momentum $\spx=\hbar\skx$. Another viewpoint
is that the used system of units provides $\hbar=1$. As we will
focus on the momentum representation, the following fact should be
mentioned. In particular, the considered approach deals with a
linear equation which contains momentum derivatives of every
order. On the other hand, the nonlinear equation of \cite{lrud16}
involves only second order spatial derivatives. The method of
\cite{lrud16} also gives a new perspective of links between
spacetime symmetries and quantum linearity \cite{parw05}.

In the one-dimensional case, the position and momentum operators
obey the deformed commutation relation \cite{kempf95}. We rewrite
the commutation relation as
\begin{equation}
\bigl[\sax,\skx\bigr]=\itt\bigl(\pen+\beta\skx^{2}\bigr)
\, . \label{gcomr}
\end{equation}
Here, the positive parameter $\beta$ is assumed to be rescaled by
factor $\hbar^{2}$ from its usual sense, and $\pen$ is the
identity operator. In the limit $\beta\to0$, the formula
(\ref{gcomr}) gives the well-known commutation relation of
ordinary quantum mechanics. Due to the Robertson formulation
\cite{robert}, the standard deviations in the prepared state
$\bro$ obey
\begin{equation}
\Delta\amh\,\Delta\bmh\geq
\Bigl|\frac{1}{2}\>\bigl\langle[\amh,\bmh]\bigr\rangle_{\bro}\Bigr|
\, . \label{robfor}
\end{equation}
By $\langle\amh\rangle_{\bro}=\Tr(\amh\,\bro)$, we mean the
quantum-mechanical expectation value. The authors of \cite{mbp16}
recently examined state-dependent uncertainty relations that are
tighter than the Roberson--Schr\"{o}dinger uncertainty relation.
Combining (\ref{gcomr}) with (\ref{robfor}) then leads to the
inequality
\begin{equation}
\Delta\sax\,\Delta\skx\geq
\frac{1}{2}
\,\bigl(1+\beta\langle\skx^{2}\rangle_{\bro}\bigr)
\geq\frac{1}{2}
\,\bigl(1+\beta(\Delta\skx)^{2}\,\bigr)
\, . \label{robr}
\end{equation}
The principal parameter $\beta$ is positive and independent of
$\Delta\sax$ and $\Delta\skx$ \cite{kempf95}. It follows from
(\ref{robr}) that $\Delta\sax$ does not exceed the square root
of $\beta$.

Following \cite{pedram12}, we will use the auxiliary wavenumber
operator $\sqx$. Let $\sax$ and $\sqx$ be self-adjoint operators
that obey $[\sax,\sqx]=\itt\pen$. In the $q$-space, the action of
$\sqx$ results in multiplying a wave function $\varphi(q)$ by $q$,
whereas $\sax\,\varphi(q)=\itt\,\xdif\varphi/\xdif{q}$. The author
of \cite{pedram12} proposed the representation
\begin{equation}
\skx=\frac{1}{\sqrt{\beta}}\>\tan\bigl(\sqrt{\beta}\sqx\bigr)
\, . \label{momr}
\end{equation}
The auxiliary wavenumber satisfies the ordinary commutation
relation but ranges between
$\pm\,q_{0}(\beta)=\pm\,\pi/(2\sqrt{\beta}\,)$. The function
$q\mapsto{k}=\tan(\sqrt{\beta}q)\big/\sqrt{\beta}$
gives a one-to-one correspondence between $q\in(-\,q_{0};+\,q_{0})$
and $k\in(-\,\infty;+\,\infty)$. So, the eigenvalues of $\skx$
fully cover the real axis. The above representation is
formally self-adjoint. However, one provides only
$\cld(\sax)\subset\cld(\sax^{\dagger})$, though
$\cld(\skx)=\cld(\skx^{\dagger})$ due to von Neumann's theorem
\cite{pedram12}.

For a pure state, we actually have three wave functions
$\phi(k)$, $\varphi(q)$, and $\psi(x)$. The auxiliary wave
function $\varphi(q)$ provides a convenient mathematical tool as
connected with $\psi(x)$ via the Fourier transform. Let the
eigenkets $|q\rangle$ of $\sqx$ be normalized through Dirac's
delta function and obey the completeness relation
\begin{equation}
\int_{-q_{0}}^{+q_{0}}
\xdif{q}\>|q\rangle\langle{q}|=\pen
\, . \label{cmr1}
\end{equation}
In the $q$-space, the eigenfunctions of $\sax$ appear as
$\langle{q}|x\rangle=\exp(-\itt{q}x)\big/\sqrt{2\pi}$. Combining
this with (\ref{cmr1}), any wave function in the coordinate space
is expressed as
\begin{equation}
\psi(x)=\frac{1}{\sqrt{2\pi}}\int_{-q_{0}}^{+q_{0}}
\exp(+\itt{q}x)\,\varphi(q)\,\xdif{q}
\, . \label{psxdf}
\end{equation}
Wave functions in the $q$- and $x$-spaces are connected by
the Fourier transform \cite{pedram12},
\begin{equation}
\varphi(q)=\frac{1}{\sqrt{2\pi}}\int_{-\infty}^{+\infty}
\exp(-\itt{q}x)\,\psi(x)\,\xdif{x}
\, . \label{phpdf}
\end{equation}
The only distinction from ordinary quantum mechanics is that each
wave function $\varphi(q)$ in the $q$-space should be formally
treated as $0$ for all $|q|>q_{0}(\beta)$.

Using the above connection, the author of \cite{pedram16}
claimed the following. The uncertainty relation of Beckner
\cite{beck} and of Bia{\l}ynicki-Birula and Mycielski
\cite{birula1} is still valid for the generalized uncertainty
principle. However, wave functions in the $q$-space play only
auxiliary role. In the GUP case, the physically legitimate
wavenumber and momentum involved in the relation (\ref{gcomr}) are
described by wavefunctions in the $k$-space. An actual
distribution of physical wavenumber values is therefore determined
with respect to $\phi(k)$ instead of $\varphi(q)$.
Let us consider the probability that momentum lies between two
prescribed values. Due to the one-to-one correspondence between
$k$ and $q$, there is a bijection between the intervals
$(k_{1};k_{2})$ and $(q_{1};q_{2})$. Thus, the probability can be
expressed as
\begin{equation}
\int_{k_{1}}^{k_{2}} |\phi(k)|^{2}\,\xdif{k}=
\int_{q_{1}}^{q_{2}} |\varphi(q)|^{2}\,\xdif{q}
\, , \label{twopr}
\end{equation}
whence $|\phi(k)|^{2}\,\xdif{k}=|\varphi(q)|^{2}\,\xdif{q}$. More
generally, two probability density functions $u(k)$ and $v(q)$ are
connected as $u(k)\,\xdif{k}=v(q)\,\xdif{q}$, in another form
\begin{equation}
u(k)=\frac{v(q)}{1+\beta{k}^{2}}
\, . \label{upvq}
\end{equation}
For pure states, when $u(k)=|\phi(k)|^{2}$ and
$v(q)=|\varphi(q)|^{2}$, the formula (\ref{upvq}) is obvious. It
is directly extended to mixed states by the spectral
decomposition.

In reality, we do not deal with the probability density functions
$u(k)$ and $w(x)$ immediately. Eigenkets of unbounded operators,
say $|k\rangle$ and $|x\rangle$, are not elements of the Hilbert
space \cite{mbg02}. Instead, we may deal with narrow distributions
that are of a finite but small width. Here, a finiteness of
detector resolution should be addressed \cite{paban13,rastcon16}.
Measuring or preparing a state with the particular value $\xi$ of
position, one is affected by some vicinity of $\xi$. In this way,
we refer to generalized quantum measurements \cite{WM10}.
Unsharpness of such measurements in the context of entropic
uncertainty relations was studied in \cite{baek16}. Dealing with a
finite-resolution measurement of the legitimate wavenumber, the
set $\clk=\bigl\{|k\rangle\langle{k}|\bigr\}$ is replaced with
some set $\clm$ of operators of the form
\begin{equation}
\mmh(\zeta):=\int_{-\infty}^{+\infty}\xdif{k}\,f(\zeta-k)\,|k\rangle\langle{k}|
\, . \label{mmzt}
\end{equation}
An acceptance function $\zeta\mapsto{f}(\zeta)$ obeys the
normalization condition, so that
$\int_{-\infty}^{+\infty}|f(\zeta)|^{2}\,\xdif\zeta=1$. Then
operators of the form (\ref{mmzt}) lead to a non-projective
resolution of the identity, namely
\begin{equation}
\int_{-\infty}^{+\infty}\xdif\zeta\,\mmh(\zeta)^{\dagger}\mmh(\zeta)=\pen
\, . \label{csrl}
\end{equation}
For the input $\bro$, the measurement results in the probability
density function
\begin{equation}
U(\zeta)=\Tr\bigl(\mmh(\zeta)^{\dagger}\mmh(\zeta)\,\bro\bigr)
=\int_{-\infty}^{+\infty}|f(\zeta-k)|^{2}\,u(k)\,\xdif{k}
\, , \label{bigu}
\end{equation}
dealt with instead of $u(k)$. If the acceptance function is sufficiently
narrow, we will obtain a good ``footprint'' of $u(k)$. Let
$\xi\mapsto{g}(\xi)$ be another acceptance function that also
obeys the normalization condition. A finite-resolution measurement
of the position is described by the set $\cln$ of operators
\begin{equation}
\nnh(\xi):=\int_{-\infty}^{+\infty}\xdif{x}\,g(\xi-x)\,|x\rangle\langle{x}|
\, . \label{nnxi}
\end{equation}
Here, the projective resolution
$\clx=\bigl\{|x\rangle\langle{x}|\bigr\}$ is replaced with
$\cln=\bigl\{\nnh(\xi)\bigr\}$. Instead
of $w(x)$, we actually deal with the probability density function
\begin{equation}
W(\xi)=\Tr\bigl(\nnh(\xi)^{\dagger}\nnh(\xi)\,\bro\bigr)
=\int_{-\infty}^{+\infty}|g(\xi-x)|^{2}\,w(x)\,\xdif{x}
\, . \label{bigw}
\end{equation}
For good acceptance functions, a distortion of
statistics will be small. The Gaussian distribution is a
typical form of such functions \cite{paban13}. It is
natural to assume that a behavior of acceptance functions is
qualitatively similar.

\section{A state-dependent bound on the sum of Shannon entropies}\label{sec3}

In this section, we will pose the generalized uncertainty
principle into a lower bound on the sum of Shannon entropies. The
usual lower bound is shown to be added by a state-dependent
correction term. We begin with differential entropies of the
Shannon type. For the given pre-measurement state $\bro$, the
wavenumber is distributed according to the probability density
function $u(k)=\langle{k}|\bro|k\rangle$, where
$k\in(-\,\infty;+\,\infty)$ and the eigenkets $|k\rangle$ are
normalized through Dirac's delta function. These kets form a projective
resolution $\clk=\bigl\{|k\rangle\langle{k}|\bigr\}$ of the
identity. Then the differential entropy is defined as
\begin{equation}
H_{1}(\clk|\bro):=-\int_{-\infty}^{+\infty} u(k)\,\ln{u}(k)\,\xdif{k}
\, . \label{denu}
\end{equation}
Similarly, we determine entropies for other continuous variables
of interest. According to (\ref{cmr1}), the eigenkets $|q\rangle$
form another resolution,
$\clq=\bigl\{|q\rangle\langle{q}|\bigr\}$, whence we write
$v(q)=\langle{q}|\bro|q\rangle$ and
\begin{equation}
H_{1}(\clq|\bro):=-\int_{-q_{0}}^{+q_{0}} v(q)\,\ln{v}(q)\,\xdif{q}
\, . \label{denv}
\end{equation}
The measurement of position is specified by the resolution
$\clx=\bigl\{|x\rangle\langle{x}|\bigr\}$. With the probability
density function $w(x)=\langle{x}|\bro|x\rangle$, one gets the
entropy $H_{1}(\clx|\bro)$.

Using (\ref{upvq}), we get the link between (\ref{denu}) and
(\ref{denv}), namely
\begin{align}
H_{1}(\clk|\bro)
&=-\int_{-q_{0}}^{+q_{0}} v(q)\,\ln{v}(q)\,\xdif{q}\,
+\int_{-\infty}^{+\infty} u(k)\,\ln\bigl(1+\beta{k}^{2}\bigr)\,\xdif{k}
\nonumber\\
&=H_{1}(\clq|\bro)+
\bigl\langle\ln(\pen+\beta\skx^{2})\bigr\rangle_{\bro}
\, . \label{hkhq0}
\end{align}
Concerning entropic uncertainty relations in the GUP case, the
following fact was noticed \cite{pedram16}. Due to (\ref{psxdf})
and (\ref{phpdf}), wave functions in the auxiliary $q$-space are
connected with coordinate wave functions via the Fourier
transform. In this regard, one merely restricts a consideration to
those functions $\varphi(q)$ that are zero beyond the range
$q\in(-\,q_{0};+\,q_{0})$. Hence, we can apply the entropic
uncertainty relation of Beckner \cite{beck} and of
Bia{\l}ynicki-Birula and Mycielski \cite{birula1},
\begin{equation}
H_{1}(\clq|\bro)+H_{1}(\clx|\bro)\geq\ln(e\pi)
\, . \label{bbmen}
\end{equation}
Although the bound (\ref{bbmen}) was first formulated for pure
states, its extension to impure ones is not difficult. The author
of \cite{pedram16} also calculated entropies of (\ref{bbmen}) for
stationary states of the harmonic oscillator in the presence of a
minimal length. In arbitrary dimensions, the energy eigenvalues
and eigenfunctions of the harmonic oscillator with the modified
commutation relation were found in \cite{cmot02}.

We already mentioned that the legitimate momentum of the
commutation relation (\ref{gcomr}) appears as $\hbar\skx$. The
wavenumber operator $\sqx$ is only a useful auxiliary tool.
Instead of $H_{1}(\clq|\bro)$, the Shannon entropy $H_{1}(\clk|\bro)$
should be used to quantify a momentum uncertainty in the presence
of a minimal observable length. Combining (\ref{hkhq0}) with
(\ref{bbmen}) gives the basic result of this section,
\begin{equation}
H_{1}(\clk|\bro)+H_{1}(\clx|\bro)\geq\ln(e\pi)
+\bigl\langle\ln(\pen+\beta\skx^{2})\bigr\rangle_{\bro}
\, . \label{bbmen1}
\end{equation}
The second quantity in the right-hand side of (\ref{bbmen1}) is a
correction of lower entropic bound in the GUP case. It is similar
to the correction term appearing in Robertson's formulation
(\ref{robr}).

In practice, we deal with the probability densities (\ref{bigu})
and (\ref{bigw}) after masking by acceptance functions. Substituting each
of these densities into the right-hand side of (\ref{denu}), we
obtain the differential entropies $H_{1}(\clm|\bro)$ and
$H_{1}(\cln|\bro)$. The entropic bound with the correction term
remains valid for these entropies, i.e.,
\begin{equation}
H_{1}(\clm|\bro)+H_{1}(\cln|\bro)\geq\ln(e\pi)
+\bigl\langle\ln(\pen+\beta\skx^{2})\bigr\rangle_{\bro}
\, . \label{bbmen11}
\end{equation}
We can prove (\ref{bbmen11}) by means of one result for integral mean
values with a weight function (see theorem 204 of the book
\cite{hardy}). Let the weight function $\lambda(x)$ be normalized.
If $\Phi^{\prime\prime}(t)$ is positive for all $t$ between
$\inf{w}(x)$ and $\sup{w}(x)$, then
\begin{equation}
\Phi\!\left(
\int \lambda(x)\,w(x)\,\xdif{x}
\right)
\leq
\int \lambda(x)\,\Phi\bigl(w(x)\bigr)\,\xdif{x}
\, . \label{thrm204}
\end{equation}
This result needs a lot of technical conditions, which are all
fulfilled in our case. If $\Phi^{\prime\prime}(t)$ is negative,
then the inequality (\ref{thrm204}) should be rewritten in
opposite direction. Combining concavity of the function
$t\mapsto-\,t\ln{t}$ with properties of acceptance functions
finally leads to the relations \cite{rastcon16}
\begin{equation}
H_{1}(\clm|\bro)\geq{H}_{1}(\clk|\bro)
\, , \qquad
H_{1}(\cln|\bro)\geq{H}_{1}(\clx|\bro)
\, . \label{nored}
\end{equation}
It is physically natural that an additional masking of acceptance
functions cannot reduce the amount of uncertainty. Combining (\ref{bbmen1}) with (\ref{nored}) at once
gives the claim (\ref{bbmen11}).

The results (\ref{bbmen1}) and (\ref{bbmen11}) lead to entropic
uncertainty relations with binning. Using some discretization, we
will always have positive entropic functions and take into account
typical experimental settings. In the case of position
measurements, values $x_{j}$ mark the ends of intervals
$\delta{x}_{j}=x_{j+1}-x_{j}$. We now deal with probabilities
\begin{equation}
p_{j}^{(\delta)}:=\int\nolimits_{x_{j}}^{x_{j+1}}
w(x)\,\xdif{x}
\, , \label{pdwx}
\end{equation}
which give the discrete distribution
$p_{\clx}^{(\delta)}$ with the Shannon entropy
$H_{1}(p_{\clx}^{(\delta)}|\bro)$. For each $j$, we apply the
above theorem for integral means, where integrals are taken over
the range between $x_{j}$ and $x_{j+1}$ and
$\lambda=1/\delta{x}_{j}$. As the function $t\mapsto-\,t\ln{t}$
is concave, the inequality (\ref{thrm204}) should be rewritten in
opposite direction. By doing some algebra, we have
\begin{equation}
{}-p_{j}^{(\delta)}\ln{p}_{j}^{(\delta)}\geq
-\int\nolimits_{x_{j}}^{x_{j+1}} w(x)\,\ln{w}(x)\,\xdif{x}
-p_{j}^{(\delta)}\ln\delta{x}_{j}
\, . \label{intres}
\end{equation}
Using $-\ln\delta{x}_{j}\geq-\ln\delta{x}$ with
$\delta{x}=\max\delta{x}_{j}$, we sum (\ref{intres}) with respect
to $j$ and get
\begin{equation}
H_{1}(p_{\clx}^{(\delta)}|\bro)\geq{H}_{1}(\clx|\bro)-\ln\delta{x}
\, . \label{hhpdx}
\end{equation}
By a parallel argument,
$H_{1}(p_{\clk}^{(\delta)}|\bro)\geq{H}_{1}(\clk|\bro)-\ln\delta{k}$,
where $\delta{k}$ is the maximal size of wavenumber bins.
Combining these inequalities with (\ref{bbmen1}) gives
\begin{equation}
H_{1}(p_{\clk}^{(\delta)}|\bro)+H_{1}(p_{\clx}^{(\delta)}|\bro)
\geq\ln\!\left(\frac{e\pi}{\delta{k}\,\delta{x}}\right)
+\bigl\langle\ln(\pen+\beta\skx^{2})\bigr\rangle_{\bro}
\, . \label{bbmen1b}
\end{equation}
Converting (\ref{bbmen11}) into uncertainty relations with binning
is obvious. We refrain from presenting the details here. The first
summand in the right-hand side of (\ref{bbmen1b}) appears just as
in entropic uncertainty relations proved in \cite{IBB84}. The
second summand reflects the presence of a minimal length.

From the physical viewpoint, the parameter $\beta$ is assumed to
be small. Taking the linear order in $\beta$, we have
\begin{align}
\bigl\langle\ln(\pen+\beta\skx^{2})\bigr\rangle_{\bro}=
\beta\langle\skx^{2}\rangle_{\bro}+O(\beta^{2})
\, . \label{betsq}
\end{align}
Thus, the correction term to the entropic bound is the doubled
correction term of Robertson's formulation (\ref{robr}). These
correction terms inspired by the GUP are physically significant
together with each other. They are always nonzero for wave
packets of a finite width. Further, the correction to the
entropic bound is bounded from above by the logarithm of the
doubled Robertson bound. For any quantum state, we have
\begin{equation}
\bigl\langle\ln(\pen+\beta\skx^{2})\bigr\rangle_{\bro}
\leq
\ln\bigl(1+\beta\langle\skx^{2}\rangle_{\bro}\bigr)
\, . \label{lrave}
\end{equation}
It can be proved on the base of the theorem (\ref{thrm204})
applied properly. Of course, any upper bound on the correction
term is not so interesting. Nevertheless, the result (\ref{lrave})
characterizes a scale in which the correction term may vary.
Moreover, the two sides of (\ref{lrave}) coincide in the term
$\sim\beta$ too.

\section{R\'{e}nyi formulation of entropic uncertainty relations}\label{sec4}

In this section, we formulate state-independent uncertainty bounds
in terms of R\'{e}nyi entropies and, with binning, in terms of
Tsallis entropies. We aim to take into account the presence of a
minimal length together with an alteration of statistics due to a
finite resolution of the measurements. When acceptance functions
of measurement apparatuses are sufficiently spread, the existing
entropic lower bounds can be improved.

For strictly positive $\alpha\neq1$, the R\'{e}nyi generalization
of $H_{1}(\clx|\bro)$ is written as
\begin{equation}
R_{\alpha}(\clx|\bro):=\frac{1}{1-\alpha}{\>}
\ln\!\left(\int\nolimits_{-\infty}^{+\infty} w(x)^{\alpha}\,\xdif{x}\right)
 . \label{recon}
\end{equation}
The standard differential entropy $H_{1}(\clx|\bro)$ is obtained
in the limit $\alpha\to1$. It will be convenient to use the
quantity
\begin{equation}
\|w\|_{\alpha}=
\left(
\int\nolimits_{-\infty}^{+\infty} w(x)^{\alpha}\,\xdif{x}
\right)^{\!1/\alpha}
 , \label{alpku}
\end{equation}
where $\alpha>0$ and $w(x)$ is positive valued. It is similar to
the definition of some norms, but gives a legitimate one only for
$\alpha\geq1$. The R\'{e}nyi entropies of discrete probability
distributions are more conventional \cite{renyi61}. For the
distribution with probabilities (\ref{pdwx}), its R\'{e}nyi
$\alpha$-entropy is defined as
\begin{equation}
R_{\alpha}(p_{\clx}^{(\delta)}|\bro):=
\frac{\alpha}{1-\alpha}{\>}\ln\bigl\|p_{\clx}^{(\delta)}\bigr\|_{\alpha}
\, , \label{rpdf}
\end{equation}
where the discrete counterpart of (\ref{alpku}) reads
\begin{equation}
\bigl\|p_{\clx}^{(\delta)}\bigr\|_{\alpha}=
\left(
\sum\nolimits_{j}\,\bigl[p_{j}^{(\delta)}\bigr]^{\alpha}
\right)^{1/\alpha}
 . \label{alpkud1}
\end{equation}
Entropic uncertainty relations with binning will be derived by
means of entropies of the form (\ref{rpdf}). The Tsallis entropies
\cite{tsallis} form another especially important family of
generalized entropies. In the discrete case, one defines
\begin{equation}
H_{\alpha}(p_{\clx}^{(\delta)}|\bro):=
\frac{1}{1-\alpha}
\left(
\bigl\|p_{\clx}^{(\delta)}\bigr\|_{\alpha}^{\alpha}-1
\right)
\, , \label{tspdf}
\end{equation}
where $0<\alpha\neq1$. In a similar way, the differential Tsallis
entropy is defined via of quantities of the form (\ref{alpku}).
For certain reasons, we will obtain Tsallis-entropy uncertainty
relations only with binning. Basic properties of generalized
entropies with application to quantum physics are considered in
\cite{bengtsson}.

As the functions $\varphi(q)$ and $\psi(x)$ are connected by
(\ref{psxdf}) and (\ref{phpdf}), certain norms of them obey
Beckner's inequalities \cite{beck}. Let positive parameters
$\alpha$ and $\gamma$ satisfy the condition $1/\alpha+1/\gamma=2$.
For $\alpha>1>\gamma$, we have the inequality
\begin{equation}
\|v\|_{\alpha}
\leq\left(\frac{1}{\varkappa\pi}\right)^{(1-\gamma)/\gamma}\|w\|_{\gamma}
\, , \label{vwinf}
\end{equation}
and its ``twin'' with swapped $v$ and $w$ \cite{barper15}. Here,
the square of $\varkappa$ is expressed as
\begin{equation}
\varkappa^{2}=\alpha^{1/(\alpha-1)}\gamma^{1/(\gamma-1)}
\, . \label{mpmpf}
\end{equation}
For a pure state, the above probability density functions read
$v(q)=|\varphi(q)|^{2}$ and $w(x)=|\psi(x)|^{2}$. Inequalities of
the form (\ref{vwinf}) are first derived for pure states and then
easily extended to impure ones \cite{barper15,IBB06}.

Using (\ref{vwinf}) and its twin, we could restate known
uncertainty relations in terms of generalized entropies. These
relations remain unchanged, only if norms in the momentum space
are calculated with the density $v(q)$. However, the probability
density function $u(k)$ is primary in the GUP case. As was
discussed above, this distribution is additionally masked with
some acceptance function. Turning (\ref{vwinf}) into
inequalities for actually registered densities $U(\zeta)$ and
$W(\xi)$, we will take into account both aspects of the problem.
Combining (\ref{bigu}) with $u(k)\,\xdif{k}=v(q)\,\xdif{q}$ leads
to
\begin{equation}
U(\zeta)=\int_{-q_{0}}^{+q_{0}}
\mu(\zeta,q)\,v(q)\,\xdif{q}
\, , \label{Uvsv}
\end{equation}
where $\mu(\zeta,q)=\bigl|f\bigl(\zeta-k(q)\bigr)\bigr|^{2}$. The
latter is sub-normalized with respect to $q$ in the sense of
\begin{equation}
J(\zeta)=\int_{-q_{0}}^{+q_{0}}\mu(\zeta,q)\,\xdif{q}=
\int_{-\infty}^{+\infty}\frac{|f(\zeta-k)|^{2}}{1+\beta{k}^{2}}\>\xdif{k}\leq
\int_{-\infty}^{+\infty}|f(\zeta-k)\bigr|^{2}\,\xdif{k}
\, , \label{Uvsv2}
\end{equation}
i.e., $J(\zeta)\leq1$. To relate $\|U\|_{\alpha}$ and $\|v\|_{\alpha}$, we use
(\ref{thrm204}) with the normalized weight
$\lambda(q)=J(\zeta)^{-1}\,\mu(\zeta,q)$. For $\alpha>1$, the
function $t\mapsto{t}^{\alpha}$ has positive second derivative.
Then the theorem (\ref{thrm204}) leads to
\begin{equation}
J(\zeta)^{1-\alpha}\,U(\zeta)^{\alpha}\leq
\int_{-q_{0}}^{+q_{0}}
\bigl|f\bigl(\zeta-k(q)\bigr)\bigr|^{2}\,v(q)^{\alpha}\,\xdif{q}
\, . \label{Uvsv3}
\end{equation}
Let us introduce the quantity
\begin{equation}
S_{f}:=\underset{\zeta}{\sup}\,J(\zeta)
=\underset{\zeta}{\sup}\int_{-\infty}^{+\infty}\frac{|f(\zeta-k)|^{2}}{1+\beta{k}^{2}}\>\xdif{k}
\, . \label{sfdef}
\end{equation}
When $\alpha>1$, we have
$S_{f}^{1-\alpha}\leq{J}(\zeta)^{1-\alpha}$ for all $\zeta$.
Substituting the latter into (\ref{Uvsv3}) and integrating over
$\zeta$, we finally get
\begin{equation}
S_{f}^{(1-\alpha)/\alpha}\,\|U\|_{\alpha}\leq
\|v\|_{\alpha}
\, . \label{Uvsv4}
\end{equation}
The probability density functions $W(\xi)$ and $w(x)$ are
connected by (\ref{bigw}). For $0<\gamma<1$, the function
$t\mapsto{t}^{\gamma}$ has negative second derivative. By a
parallel argument, we obtain $\|w\|_{\gamma}\leq\|W\|_{\gamma}$.
It then follows from (\ref{vwinf}) that
\begin{equation}
\|U\|_{\alpha}
\leq\left(\frac{S_{f}}{\varkappa\pi}\right)^{(1-\gamma)/\gamma}\|W\|_{\gamma}
\, , \label{UWinf}
\end{equation}
where $1/\alpha+1/\gamma=2$ and $\alpha>1>\gamma$. At the last
step, we take into account that
$(1-\alpha)/\alpha=(\gamma-1)/\gamma$. By a parallel argument, the
inequality between $\|w\|_{\alpha}$ and $\|v\|_{\gamma}$ is
transformed into
\begin{equation}
\|W\|_{\alpha}
\leq\left(\frac{S_{f}}{\varkappa\pi}\right)^{(1-\gamma)/\gamma}\|U\|_{\gamma}
\, , \label{WUinf}
\end{equation}
under the same conditions on $\alpha$ and $\gamma$. The formulas
(\ref{UWinf}) and (\ref{WUinf}) immediately lead to uncertainty
relations in terms of R\'{e}nyi entropies. To do so, we take the
logarithm of (\ref{UWinf}) and (\ref{WUinf}) and then use the
expression of entropies via $\|U\|_{\alpha}$ and so on. In the
presence of a minimal observable length, we obtain
\begin{equation}
R_{\alpha}(\clm|\bro)+R_{\gamma}(\cln|\bro)
\geq\ln\!\left(\frac{\varkappa\pi}{S_{f}}\right)
 , \label{remnag}
\end{equation}
where $1/\alpha+1/\gamma=2$ and $S_{f}$ is defined by
(\ref{sfdef}). The presented entropic bound is independent of the
measured state. For the fixed entropic orders $\alpha$ and
$\gamma$, the right-hand side of (\ref{remnag}) depends only on
$S_{f}$, i.e., on the momentum acceptance function and the
minimal-length parameter $\beta$. As follows from (\ref{mpmpf}),
the term $\varkappa$ increases from $\varkappa=2$ for
$\gamma=1/2$ up to $\varkappa=e$ for $\gamma=1$. For the
particular choice $\alpha=\gamma=1$, we therefore have
\begin{equation}
H_{1}(\clm|\bro)+H_{1}(\cln|\bro)
\geq\ln\!\left(\frac{e\pi}{S_{f}}\right)
 . \label{shamnag}
\end{equation}

Let us take the probability distributions $p_{\clm}^{(\delta)}$
and $p_{\cln}^{(\delta)}$ obtained by discretization of the
densities $U(\zeta)$ and $W(\xi)$ with respect to chosen marks on
the $\zeta$- and $\xi$-axes. Probabilities are determined by
integrals similarly to (\ref{pdwx}). Assuming
$1/\alpha+1/\gamma=2$ and $\alpha>1>\gamma$, one can convert
(\ref{UWinf}) and (\ref{WUinf}) into the inequality
\begin{equation}
\bigl\|p_{\clm}^{(\delta)}\bigr\|_{\alpha}
\leq\left(\frac{S_{f}\,\delta\zeta\,\delta\xi}{\varkappa\pi}\right)^{(1-\gamma)/\gamma}
\bigl\|p_{\cln}^{(\delta)}\bigr\|_{\gamma}
\, , \label{bUWinf}
\end{equation}
and its ``twin'' with swapped $p_{\clm}^{(\delta)}$ and
$p_{\cln}^{(\delta)}$. For $1/\alpha+1/\gamma=2$, we finally
obtain entropic uncertainty relations with binning,
\begin{equation}
R_{\alpha}(p_{\clm}^{(\delta)}|\bro)+R_{\gamma}(p_{\cln}^{(\delta)}|\bro)
\geq\ln\!\left(\frac{\varkappa\pi}{S_{f}\,\delta\zeta\,\delta\xi}\right)
 . \label{bremnag}
\end{equation}

To derive uncertainty relations in terms of R\'{e}nyi entropies,
purely algebraic operations were used. The case of Tsallis
entropies is not so immediate. We will adopt the method of
\cite{rast104}, where the minimization problem was examined. It is
essential for optimization that norm-like functionals of discrete
probability distributions obey inequalities of the form
\begin{equation}
\bigl\|p_{\clm}^{(\delta)}\bigr\|_{\alpha}\leq1\leq
\bigl\|p_{\clm}^{(\delta)}\bigr\|_{\gamma}
\, , \label{papb1}
\end{equation}
where $\alpha>1>\gamma$. For probability density functions, such
inequalities do not hold always. Hence, we will
consider Tsallis entropies only with binning. For $0<\alpha\neq1$
and $y>0$, we put the $\alpha$-logarithm
$\ln_{\alpha}(y):=\bigl(y^{1-\alpha}-1\bigr)/(1-\alpha)$.
The Tsallis entropies of the corresponding discrete distributions
satisfy
\begin{equation}
H_{\alpha}(p_{\clm}^{(\delta)}|\bro)+H_{\gamma}(p_{\cln}^{(\delta)}|\bro)
\geq\ln_{\nu}\!\left(\frac{\varkappa\pi}{S_{f}\,\delta\zeta\,\delta\xi}\right)
 , \label{btsmnag}
\end{equation}
where $1/\alpha+1/\gamma=2$ and $\nu=\max\{\alpha,\gamma\}$. It
follows from (\ref{Uvsv2}) that $S_{f}\leq1$. Hence, the
uncertainty relations (\ref{remnag}), (\ref{bremnag}), and
(\ref{btsmnag}) remain valid with $1$ instead of $S_{f}$.

The presented reasons also give an evidence that the existence of
a minimal length will increase state-independent entropic
bounds. This increasing takes place, when the quantity
(\ref{sfdef}) is strictly less than $1$. As an example,
we consider the momentum acceptance function
\begin{equation}
|f(\zeta)|^{2}=\frac{1}{\sigma\sqrt{2\pi}}
\>\exp\!\left(-\frac{\zeta^{2}}{2\sigma^{2}}\right)
 . \label{gaufun}
\end{equation}
Substituting (\ref{gaufun}) into (\ref{Uvsv2}) gives
\begin{equation}
S_{f}\leq\frac{1}{\sigma\sqrt{2\pi}}
\,\int_{-\infty}^{+\infty}\frac{\xdif{k}}{1+\beta{k}^{2}}
=\sqrt{\frac{\pi}{2\sigma^{2}\beta}}
\, . \label{jzgs}
\end{equation}
The above entropic relations can all be recast with the
right-hand side of (\ref{jzgs}) instead of $S_{f}$. It is strictly
less than $1$, whenever one has $\sigma>\sqrt{\pi/(2\beta)}$. If the
momentum acceptance function is not sufficiently narrow, we will
deal with increasing of state-independent lower bounds in entropic
uncertainty relations. Addition amount of uncertainty may be due
to limitations of measurement resolution. It seems that more
accurate bounds on (\ref{sfdef}) could be obtained. This question
could be a subject of separate investigation.

\section{Conclusions}\label{sec5}

We have examined the question how entropic uncertainty relations
for position and momentum are changed in the GUP case. Our
approach is based on the formally self-adjoint representation
proposed in \cite{pedram12}. To describe the momentum, two
different operators are used here. We showed that lower entropic
bounds are increased, when the physically true momentum is dealt with.
Just this momentum obeys the modified commutation relation.
Increasing of entropic bounds has been formulated in two distinct
ways. The known lower bound on the sum of differential Shannon
entropies is added by the expectation value of certain operator.
Taking the linear order in $\beta$, this correction term coincides
with the doubled correction inspired in Robertson's formulation.
We further formulated state-independent uncertainty relations in
terms of generalized entropies, both differential and with
binning. A finiteness of measurement resolution was naturally
involved into the consideration. It has been shown that entropic
lower bounds for actually measured distributions are generally
increased in the GUP case.

\end{document}